\documentstyle[sprocl,epsfig]{article}

\def\lapproxeq{\lower .7ex\hbox{$\;\stackrel{\textstyle                     
<}{\sim}\;$}}                                                               
\def\gapproxeq{\lower .7ex\hbox{$\;\stackrel{\textstyle                     
>}{\sim}\;$}}                                                               
\def\msb{\overline{\rm MS}}                                                   
\def\gup{{g\uparrow}} 
\def\gdown{{g\downarrow}} 
\def\asup{{\alpha_S\uparrow\uparrow}}                                         
\def\asdown{{\alpha_S\downarrow\downarrow}}                                   

\def\be{\begin{equation}}
\def\ee{\end{equation}}
\def\bea{\begin{eqnarray}}
\def\eea{\end{eqnarray}}

\begin{document}

\title{PARTON DISTRIBUTIONS : A NEW GLOBAL ANALYSIS\footnote{Talk
by Dick Roberts at DIS98 workshop, Brussels, April 1998}}

\author{A.D. MARTIN$^a$, R.G. ROBERTS$^b$, W.J. STIRLING$^{a,c}$, 
and R.S. THORNE$^d$}

\address{$^a$ Dept. of Physics, University of Durham, Durham, DH1 3LE\\
$^b$ Rutherford Appleton Laboratory, Chilton, Didcot, Oxon, OX11 0QX\\
$^c$ Dept. of Mathematical Sciences, University of Durham, Durham, DH1 3LE\\
$^d$ Jesus College, University of Oxford, Oxford OX1 3DW}


\maketitle\abstracts{We present a new analysis of parton distributions of the 
proton.  This incorporates a wide range of new data, an improved treatment of 
heavy flavours and a re-examination of prompt photon production.  The new set 
(MRST) shows systematic differences from previous sets of partons which can be
identified with particular features of the new data and with improvements in 
the analysis. An `explanation' of the behaviour seen on the Caldwell plot
is offered.}
 
\section{New data}

New experimental information used in the anaysis includes the following.                          
\begin{itemize}                                                               
\item[(i)]  New, more precise, measurements of the structure function         
$F_2^{ep}$ for deep inelastic electron-proton scattering by the H1 and        
ZEUS collaborations at HERA \cite{H1,ZEUS}.                                   
                                                                              
\item[(ii)] A re-analysis of the CCFR neutrino data leading to a new set of   
$F_2^{\nu N}$ and $x F_3^{\nu N}$ structure function measurements \cite{CCFR2}.
                                                                               
\item[(iii)]  Measurements of the charm component of $F_2^{ep}$                
in electron-proton deep inelastic scattering at HERA \cite{H1C,ZEUSC}.  
These complement the existing EMC charm data \cite{EMC}.                       
                                                                               
\item[(iv)]  Very precise measurements of prompt photon production, 
from the E706 collaboration \cite{E706}.
  These data motivate us to look again at our treatment of this                
reaction, and in particular of the WA70 prompt photon measurements \cite{WA70}.
We emphasize that such prompt photon data are the main constraints on the      
gluon.
                                                                                
\item[(v)] The E866 collaboration \cite{E866} have measured the asymmetry in   
Drell-Yan production in $pp$ and $pn$ collisions over an extended $x$ range,   
$0.03 \lapproxeq x \lapproxeq 0.35$.  The asymmetry data provide direct 
information                                             
on the $x$ dependence of the difference, $\bar{u} - \bar{d}$, of the           
sea quark densities.                                                           
                                                                               
\item[(vi)]  The new CDF \cite{CDF} 
data on the asymmetry of the rapidity                                          
of the charged lepton from $W^\pm \rightarrow l^\pm \nu$ decays                
at the Tevatron $p \bar{p}$ collider.  The new data extend to larger           
values of lepton rapidity.  
                                                                               
\item[(vii)]  The final NMC analysis on $F_2^{\mu p}$, $F_2^{\mu d}$ and the 
ratio $F_2^{\mu d}/F_2^{\mu p}$ \cite{NMC}.                                    
                                                                               
\item[(viii)]The data on Drell-Yan production obtained by the E772 
collaboration \cite{E772} cover a wider range of $x_F$ than the E605 
data \cite{E605} which we                                             
have used to constrain the sea.
\end{itemize}       
                                                                               
\section{Improvements in the analysis}                                         
                          
We parametrize the starting parton distributions at $Q^2 = Q^2_0               
= 1$~GeV$^2$ where the number of active flavours is $n_f=3$. We work in the    
$\rm\overline{MS}$ renormalization scheme and use the standard starting 
MRS parametric forms.                                                          
                                                                               
For the first time our treatment of the heavy flavour densities, charm and     
bottom, is on a firm theoretical footing. These densities are determined by the
other parton distributions and no extra parameters are introduced apart from 
the heavy quark masses. At very low $Q^2$ the structure functions 
$F^H_2(x,Q^2)$, with $H=c,b,$ are described by boson-gluon fusion and the heavy
quark densities turn on at $Q^2\simeq m^2_H$. The procedure is described in 
the talk by Robert Thorne\cite{robtalk}. 
                   
A special feature of our analysis is the particular attention to the 
uncertainties in the gluon distribution at large $x$. The main constraints in 
this region are data on
prompt photon production in $pp$ or $pA$ collisions from the WA70 \cite{WA70}  
and the E706 \cite{E706} experiments.            
We begin by taking a canonical value of $\left<k_T\right> = 0.4$~GeV           
for the the WA70 data. We then explore a range of                              
gluon distributions in which $\left<k_T\right>$                                
goes from one extreme of $\left<k_T\right> =0$ to the other $\left<k_T\right>
 = 0.64$~GeV. 
We call the gluon distributions which correspond to 
$\left<k_T\right>=$                                 
0, 0.4 and 0.64~GeV the higher, central and lower (large $x$) gluons 
respectively --- since a smaller gluon density is compensated by a 
larger $\left<k_T\right>$.  
We denote the corresponding parton sets by MRST($\gup$), MRST and 
MRST($\gdown$).
                                                                               
The optimum global MRST description has the QCD parameter                      
$\Lambda_{\rm\overline{MS}} (n_f=4) = 300$~MeV, which corresponds to           
$\alpha_S(M_Z^2) = 0.1175$.

The starting gluon and sea-quark distributions at small $x$, 
$xg \sim x^{-\lambda_g}$, $xS \sim x^{-\lambda_S}$ and we find 
negative values of $\lambda_g$ , which                                      
imply a `valence' type of behaviour for the gluon at low $x$ for               
$Q^2=1$~GeV$^2$.   As $Q^2$                                                    
increases the valence-like                                                     
character of the gluon rapidly disappears due to evolution being driven by the 
steeper sea, and by $Q^2 \simeq 2$~GeV$^2$ the                            
gluon is \lq flat\rq~in $x$, that is $\lambda_g = 0$.                          
                                                                               
\begin{figure}[hbt] 
\vspace{-1.2cm}                                                                
\begin{center}     
\epsfig{figure=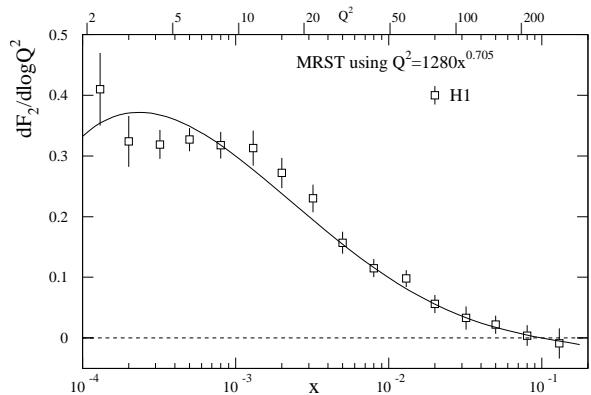,height=12cm}    
\end{center}    
\vspace{-6.0cm}                                                                
\caption{The description of the slopes $\partial F_2/\partial     
\ln Q^2$ versus $x$ for the HERA data.}    
\label{fig:slope}
\end{figure}

Fig.~\ref{fig:slope} is our version of the `Caldwell plot'\cite{ac}
showing $\partial F_2/\partial \ln Q^2$ versus $x$.  The curve 
shows the values of                                                   
the slope $\partial F_2/\partial \ln Q^2$ versus $x$ for the H1 
data, compared with the slope found in the MRST fit (evaluated at the 
particular values of $Q^2$ appropriate to the experimental data).  
Thus the claim\cite{ahm} of a sudden change of dynamics at low $x$ may 
be attributed simply to the rapidly varying
nature of the gluon at very low $Q^2$.

However, this `conventional' explanation for the Caldwell plot may, in turn,
be a signal for some breakdown of exact DGLAP around 1 GeV$^2$; certainly we
cannot move lower down in $Q^2$ as the gluon density (and hence $F_L$)
starts to become negative at very low $x$.

\section{Summary}
 
This talk is an advertisement for the MRST parton sets. The full details
of the analysis are available\cite{mrst}. In addition to the sets MRST,
MRST($\gup$), MRST($\gdown$) which explore the likely allowed variation
of the gluon we have sets MRST($\asup$) and MRST($\asdown$) which explore
a variation of $\alpha_S$ of 0.005 about the central value of 0.1175.
These are all NLO sets evaluated in the $\msb$ scheme and are available 
from the Durham database (http://durpdg.dur.ac.uk/ HEPDATA) together with
a package which computes the relevant structure functions including $F_L$
and $F_2^{charm}$. Furthermore, the corresponding sets evaluated in the DIS 
scheme can be likewise accessed.

\end{document}